\documentclass[final]{aipproc}

\layoutstyle{8x11double}

\newcommand{\apj}{{\it ApJ}}

\newcommand{\nat}{{\it Nature}}

\newcommand{\etal}{\mbox{et al.}}

\begin{document}

\title[Burst Oscillations]{Millisecond Oscillations During Thermonuclear 
X-ray Bursts}
\author{Michael P. Muno}{address={Hubble Fellow, Department of Physics and 
Astronomy, University of California, Los Angeles, CA 90095}}

\begin{abstract}
I review the basic phenomenology and theory of the millisecond brightness
oscillations observed during thermonuclear X-ray bursts from 
13 of $\approx 70$ accreting 
neutron stars in low-mass X-ray binaries. Compelling observations indicate 
that the oscillations are produced by surface brightness patterns on the 
rapidly rotating neutron stars. However, it remains to be understood 
(1) why the 
brightness patterns producing them persist for up to 15 s during an X-ray 
burst, whereas the burning should cover the 
entire surface in less than 1 s, and (2) why the frequencies drift upward 
by $\approx 5$ Hz during the course of the burst. These peculiarities can 
probably be explained by taking into account the expansion of the surface 
layers caused by the burning, zonal flows that form due to pressure gradients 
between the equator and poles, and
Rossby-Alfv\'{e}n modes that are excited in the surface ocean. Further progress
toward understanding how burning progresses on the surface of the neutron 
star can be made with a next-generation X-ray timing mission, which would 
provide a larger sample of sources with oscillations, detect
sideband signals produced by the spectrum of modes that should be excited
in the neutron star ocean, and measure 
harmonic structure in the profiles of the oscillations. 
These observations would be crucial for measuring
the distribution of the rotation rates of neutron stars, 
the progression of unstable nuclear burning in the accreted ocean, 
and the curvature of the space-time around the neutron star.
\end{abstract}

\maketitle

\section{Introduction}

Thermonuclear X-ray bursts are produced when helium on the surface of an 
accreting neutron star ignites unstably \citep[see][for a review]{sb03}. 
The accreted material itself is
usually hydrogen. The helium is produced because the accreted hydrogen is 
compressed and heated by the column of material above it, and begins to 
burn steadily via the CNO cycle. Eventually, the critical temperature and 
density are reached such that He can burn via a triple-$\alpha$ process. 
The burning is unstable, and engulfs the entire surface of the neutron 
star in less than a second. This results in a $10^{38}$ erg s$^{-1}$ flash 
of X-rays that out-shines the emission from the accretion flow for tens 
of seconds. These bursts recur on time scales of hours to days, and so 
multiple bursts have been observed from the neutron stars in $\approx 65$
low-mass X-ray binaries.

The unstable burning is likely to begin in a small region on the neutron 
star, and so it has long been expected that the resulting hot spot should 
produce modulations in the burst flux at the
spin period of the neutron star. Indeed, ``burst oscillations'' have now 
been observed with {\it RXTE} from 13 neutron star LMXBs 
(Table~\ref{tab:init}). There are several reasons to believe that 
the burst oscillations occur at the spin frequencies of the neutron 
stars. First, and most importantly, two of these sources are X-ray pulsars, 
which exhibit periodic modulations in the persistent emission between
bursts at the same frequency as the burst oscillations \citep{cha03,str03}. 
The frequencies 
of all of the oscillations are characteristic to each source, and are 
distributed uniformly between 270 and 620 Hz.
Second, once one accounts for a small frequency drift (described below) 
the burst oscillations are nearly coherent \citep{str96,sm99}. 
In one case, the oscillations
are observed to be coherent for $10^5$ cycles
during a carbon superburst \citep{sm02}. Third,
the maximum frequencies of the oscillations are stable to
within a few parts in a thousand in bursts separated by several years
\citep{str98, sm99, gil02, mun02a}. Fourth, the 
oscillations are strongest
in the rises of bursts, when the nuclear burning is likely to be confined to
small areas on the surfaces of the neutron stars \citep{str98}.
Finally, in the tails of the bursts, the amplitudes of the oscillations as a 
function of energy are consistent with those expected from temperature 
variations of $\approx 0.2$ keV across the surface of the neutron star
\citep{moc03}. As signals from the 
surfaces of neutron stars, these oscillations can be used to study the 
evolution of the spin frequencies of accreting neutron stars, the spacetime
around the star, and how 
thermonuclear burning proceeds on the stellar surface. 
\begin{figure}[thb]
\resizebox{.7\textwidth}{!}
 {\includegraphics{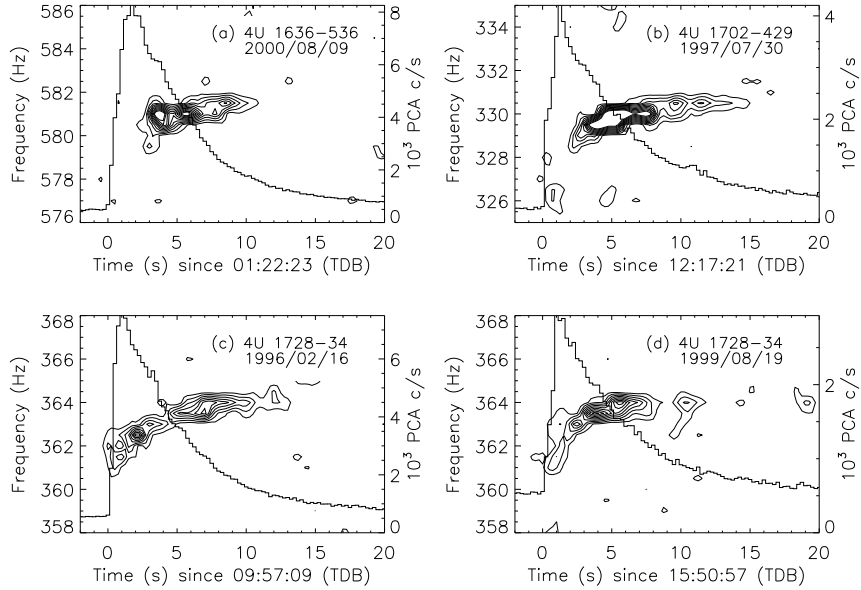}}
\caption{Dynamic power spectra illustrating the frequency
evolution of burst oscillations. Contours of power as a function of 
frequency and time were generated from power spectra of 2 s intervals
computed every 0.25 s. A Welch function was used to taper the data 
to reduce sidebands in the power spectrum due to its finite length. 
The contour levels are at powers of $0.02$ in single-trial probability 
starting at a chance occurrence of $0.02$. The PCA count rate is plotted 
referenced to the right axis.
\label{fig:freqev}}
\end{figure}

However, the simplest models of inhomogeneous burning fail to explain two 
aspects of the oscillations, which are illustrated in Figure~\ref{fig:freqev}.
First, the oscillations persist for up to 
15 s during a burst, long after the burning should have engulfed the
entire surface of the neutron star. Second, they drift upward in frequency by 
up to 5 Hz during the course of a burst, which suggests that the brightness
pattern moves opposite the sense of the rotation, such that 
$\omega_{\rm obs} = \Omega_{\rm NS} - \omega_{\rm pattern}(t)$ \citep{str97a}.
Several models have been proposed to explain one or both of these aspects of 
the oscillations.

\section{Models of the Oscillations}

The upward sense of the frequency drift could be explained by the conservation
of angular momentum in an expanding burning layer \citep{str97a}.
Under this model, the energy released in the first second of the burst 
causes the burning layer to expand and slow relative to the rotation of the 
neutron star. The frequency drift is observed as the burning layer cools 
and re-couples to the rest of the neutron star, causing the frequency of the
oscillations to increase. Unfortunately, it appears that 
too little energy is released during
a burst to cause the burning layer to expand to the height required to 
explain the observed frequency drifts \citep[compare][]{cb00,cum02}.

An additional frequency drift could be produced by 
accounting for the propagation of the cooling front after the fuel has been
exhausted \citep{slu02}. If the burst ignites 
near the rotational equator, then as the burst cools the pressure at the poles
is likely to be larger than at the equator. The pressure gradient and the
Coriolis force would then combine to generate a zonal flow opposite the 
rotation of the neutron star, in the same manner as the trade winds are 
formed on Earth. As the entire surface cools, the velocity of the flow 
should slow. Any brightness patterns 
\begin{table}
\begin{tabular}{lcc}
 \tablehead{1}{c}{b}{Source} & \tablehead{1}{c}{b}{$\nu_{\rm burst}$\\(Hz)} 
& \tablehead{1}{c}{b}{Ref.}\\
4U 1608--522 & 620 & --- \\ 
SAX~J1750$-$2900 & 600 & \citep{kaa02} \\
MXB 1743--29 & 589 & \citep{str97a} \\
4U 1636--536 & 581 & \citep{zha97} \\
MXB 1659--298 & 567 & \citep{wsf00} \\
Aql X-1 & 549 & \citep{zha98} \\
KS 1731--260 & 524 & \citep{smb97} \\ 
SAX J1748.9-2901 & 410 & \citep{kaa03} \\
SAX J1808.4--3658 & 401 & \citep{cha03} \\
4U 1728--34 & 363 & \citep{str96} \\
4U 1702-429 & 329 & \citep{sm99} \\
XTE J1814--338 & 314 & \citep{str03} \\
4U 1916--053 & 270 & \citep{gal01}
\end{tabular}
\caption{Sources of Burst Oscillations}
\label{tab:init}
\end{table}
on the surface would be carried along 
the zonal flow, again producing oscillations with a frequency lower than 
that of the neutron star's rotation. Unfortunately, neither of the above
models can easily explain the persistence of the oscillations beyond the
first second of a burst.

\begin{figure}[bt]
\resizebox{.7\textwidth}{!}
 {\includegraphics{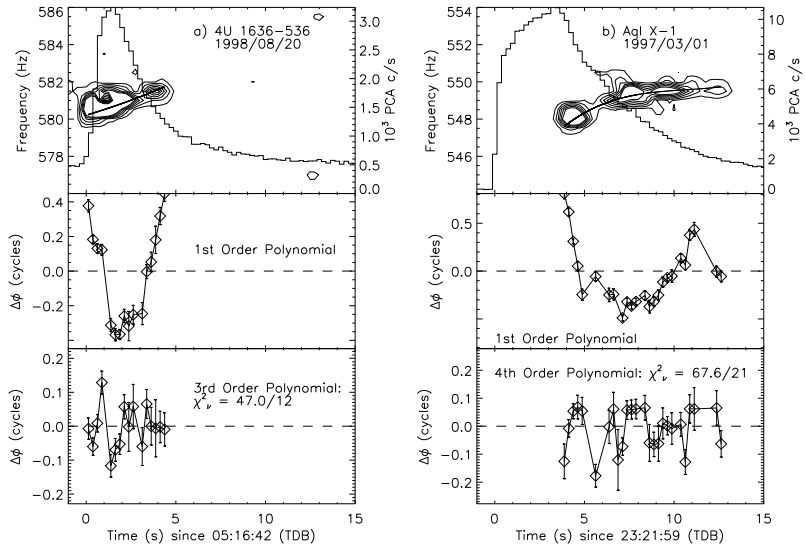}}
\caption{Two bursts for which smooth phase evolution models fail to 
reproduce the observed signals. {\it Top panel}: Dynamic power spectrum
of a burst oscillation, as in Figure~1. The thick solid line illustrates 
the best-fit frequency evolution. {\it Middle panel}: The phase residuals
assuming the oscillation trains are produced by a signal with constant 
frequency. {\it Bottom panel}: The phase residuals assuming that the
oscillations evolves according to the models indicated. Sudden 0.1 cycle
changes in phase are evident in the residuals.
\label{fig:jumps}}
\end{figure}
The unstable nuclear burning that initiates a burst may also excite 
oscillatory modes in the surface layers of the neutron star 
\citep{hey03,lee03}. The modes should propagate around the neutron star in the
non-rotational frame, with a frequency that depends on the temperature 
and composition of the surface layers. Although the frequencies 
of the most common gravity, Kelvin, and Rossby modes\footnote{Note
that these surface modes are present only in the outer $\approx 10$ m of
the neutron star, in contrast to the global modes that are invoked to 
limit the rotational frequency of the star by producing gravitational 
radiation.}
 are too large to 
explain the frequency drifts observed, Fred Lamb has suggested that 
Rossby-Alfv\'{e}n modes may have the right frequency.
However, in order to explain the 
burst oscillations, some unknown mechanism must select a single,
dominant surface mode, and that mode must always propagate with retrograde
motion.

\section{Key Observations}

To better constrain the above models for the burst oscillations, we have
carried out a systematic study of them using a phase connection technique 
commonly used on radio pulsars (Figure~\ref{fig:jumps}). 
To implement this technique, we fold
the data in short (0.25 s) intervals about a trial phase model, generally 
corresponding to a constant frequency, and then measure the phases of each 
folded profile. We then derive a correction to the initial phase models, by 
modeling the residuals from the phase model with either a polynomial or 
saturating exponential using a $\chi^2$ minimization technique. We iterate 
this process until $\chi^2$ is minimized when comparing the phase residuals 
to zero. We use the resulting phase models to 
quantify the coherence of the oscillations \citep{mun02a} and to examine 
their profiles \citep{moc02}.

\begin{figure}[t]
\resizebox{.9\columnwidth}{!}
 {\includegraphics{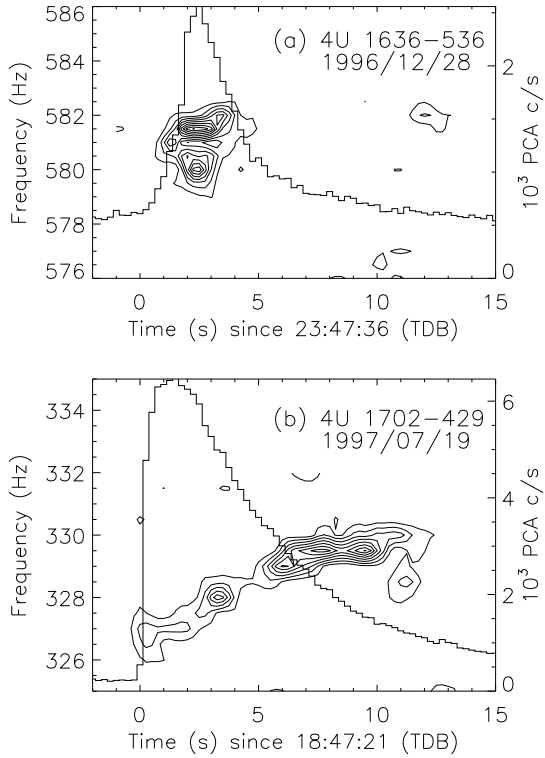}}
\caption{Same as Figure~1, for oscillations that exhibited two 
simultaneous
signals separated in frequency by $\sim 1$~Hz. The secondary signal 
occurs at 2--3~s in panel a, with a chance probability that it is due to
noise of $3\times10^{-10}$. The second signal at 11~s in panel b has a 
chance probability of $5 \times 10^{-5}$. 
\label{fig:twofreq}}
\end{figure}

If the oscillations are coherent, then a smooth frequency model should be 
able to predict their phase as a function of time. Although the oscillations
appear coherent using this measure in most cases, Figure~\ref{fig:jumps} 
illustrates two cases in which the best-fit phase model failed to produce
a $\chi^2$ that is consistent with zero phase residuals. Out of a sample
of 59 oscillation trains, 20\% did not appear to evolve smoothly in phase
\citep{str01,mun02a}. There are several possible causes of 
this, including: 0.1 cycle phase jumps that could represent
sudden changes in the position of the brightness pattern on the star; 
1 Hz s$^{-1}$ frequency shifts that could be caused by sudden changes in the 
velocity of the brightness pattern; and multiple signals present simultaneously
with frequency separations of $< 1$ Hz that otherwise cannot be resolved 
with Fourier techniques. This latter possibility is the most interesting, 
because in at least two instances out of $\approx 100$ oscillation trains,
multiple 
signals have been observed simultaneously with frequency separations of 
$\approx 1$ Hz (Figure~\ref{fig:twofreq}) \citep{mun02a, mil00, gal01}.
Moreover, Deepto Chakrabarty has reported sideband signals 30~Hz below
the main burst oscillation in a couple bursts from 4U 1728--34. 
The simultaneous presence of multiple signals would provide compelling 
evidence that the oscillations are produced by modes in the surface with 
a spectrum of latitudinal and radial wave numbers.

\begin{figure}[th]
\resizebox{.9\columnwidth}{!}
 {\includegraphics{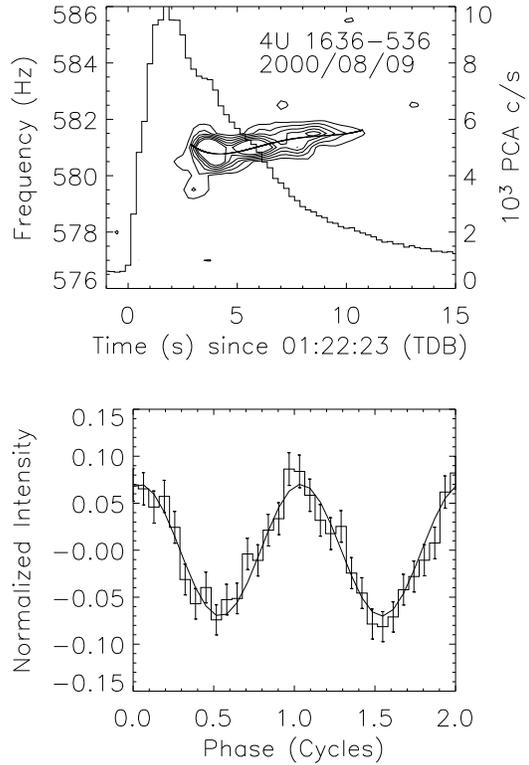}}
\caption{{\it Top panel}: Dynamic power spectrum
of a burst oscillation, as in Figure~1. The thick solid line illustrates 
the best-fit frequency evolution. {\it Bottom panel}: Profile of the 
oscillations produced by folding the data about the best-fit phase
model. This oscillation train, and all others that we have examined,
exhibit no detectable harmonic content.
\label{fig:folded}}
\end{figure}

The phase models also allow us to coherently fold the oscillation trains to 
examine their amplitudes and profiles (Figure~\ref{fig:folded}) \citep{moc02}.
The amplitude of a 
typical oscillation is 5\% rms, and they almost always appear 
sinusoidal,\footnote{The millisecond
pulsar XTE J1814--338 provides the only exception, probably because
its relatively strong magnetic field beams the emission along the normal
to the surface of the neutron star \citep{str03,bat04}.} 
with upper limits of $< 2$\% to the amplitudes of harmonic and half-frequency 
signals. Since the phases of the oscillations 
are well-measured, we can 
also coherently summed the profiles from all of the oscillations for each 
source. For the sources with the largest number of observed 
oscillation trains, 4U 1728--34 and 4U 1636-536, the summed profiles provide 
strict upper limits of 0.3\% to the amplitude of any harmonic content
(Table~\ref{tab:harm}).

We have computed the profiles 
expected from a bright region on a neutron star, including the expected
Doppler and General-relativistic light-bending effects, and find that the
only plausible ways to explain this lack of harmonic content are if (1) the 
bright region covers nearly half the neutron star, or (2) a corona of 
electrons around the neutron star scatters and attenuates the signal from
the surface \citep{moc02}. If the brightness pattern is indeed 
symmetric, then this suggests that the oscillations could be produced 
by a mode in the surface layers of the neutron star with azimuthal wave 
number $m=1$. However, the scattering hypothesis also merits further study, 
because energy-dependent scattering could explain the observed lack of the
expected Doppler signatures in the phases of the oscillations as a function
of energy \citep{for99,moc03}. 
The amount of scattering should be determined by simultaneously modeling the 
burst spectra and the energy dependence of the oscillation amplitudes and 
phases.

\begin{table}[thb]
\begin{tabular}{lcccccccc}
\tablehead{1}{c}{b}{(1)\\Source} & \tablehead{1}{c}{b}{(2)\\No. Osc.} & 
\tablehead{1}{c}{b}{(3)\\Counts} & \tablehead{1}{c}{b}{(4)\\Background} & 
\tablehead{1}{c}{b}{(5)\\$A_{1/2}$} & \tablehead{1}{c}{b}{(6)\\$A_{1}$} & 
\tablehead{1}{c}{b}{(7)\\$A_{3/2}$} & \tablehead{1}{c}{b}{(8)\\$A_{2}$} \\
4U 1636--536 & 17\tablenote{11 oscillations were used to constrain $A_{1/2}$ and $A_{3/2}$, for a total of $9.8\times10^5$ counts with $1.0\times10^5$ counts background.} & $1.1\times10^6$ & $1.3\times10^5$ & $< 0.6$ & 5.4(3) & $< 0.5$ & $< 0.3$ \\
MXB 1659--298 & 3 & $2.8\times10^4$ & $6.1\times10^3$ & $< 2.7$ & 9.3(8) & $< 2.8$ & $< 2.7$ \\
Aql X-1 & 3 & $4.2\times10^5$ & $2.0\times10^4$ & $< 0.6$ & 3.3(1) & $< 0.5$ & $< 0.5$ \\
KS 1731--260 & 4 & $2.5\times10^5$ & $4.5\times10^4$ & $< 1.2$ & 4.7(2) & $< 0.9$ & $< 0.6$ \\ 
4U 1728--34 & 24\tablenote{13 oscillations were used to constrain $A_{1/2}$ 
and $A_{3/2}$, for a total of $1.2\times10^6$ counts with $1.9\times10^5$ counts background.} & $1.6\times10^6$ & $2.3\times10^5$ & $< 0.6$ & 5.5(1) & $< 0.6$ & $< 0.3$ \\
4U 1702--429 & 8 & $6.1\times10^5$ & $5.6\times10^4$ & $< 0.6$ & 4.6(2) & $< 0.7$ & $< 0.7$ \\
\end{tabular}
\caption{Harmonic Amplitudes of Burst Oscillations. Columns are as follows: 
(1) Source name. 
(2) Number of bursts with oscillations
used to make a combined profile. (3) Total number of counts in the profile, 
including background. (4) Estimated background counts in the profile. 
(5-8) Percent fractional rms amplitudes,
or 95\% upper limits on amplitudes at $n=$0.5,1,1.5, and 2 times the main
frequency.}
\label{tab:harm}
\end{table}

\section{The Future}

Observations with {\it RXTE} have clearly established that the oscillations 
observed during thermonuclear X-ray bursts occur at the spin periods of 
the underlying neutron stars, and have provided tantalizing clues as to 
how nuclear burning proceeds on the surfaces of neutron stars. In the 
short term, it would be helpful to address outstanding theoretical 
questions, such as the expected frequencies of Rossby-Alfv\'{e}n modes, 
and the amount the oscillation signals are attenuated by scattering as they
propagate away from the neutron star. However, a future
X-ray timing mission truly is needed to bring these initial observations to 
their full potential by making several crucial observations.

First, oscillations are currently only observed from 12 of $\approx 65$ 
bursting LMXBs, and are only observed from about half of the bursts
from any given source. The difference between the amplitudes of the
detected oscillations and the upper limits to the non-detections is only
a factor of two \citep{mgc04}, so a future mission with a larger 
effective area could greatly increase the number of neutron stars with burst 
oscillations, and consequently with known spin periods. A larger sample would 
be important
for understanding the distribution of observed spin periods, and thus for 
exploring why all neutron stars appear to be rotating significantly below their
break-up frequency \citep{wz97,bil98}.

Second, if the burst oscillations are indeed produced by modes in the surface
layers, a future X-ray timing mission should detect a spectrum of signals 
with different latitudinal and radial wave numbers. The spacing of these modes
would allow us to measure the pressure, density, and composition of the
burning layers on a neutron star \citep[e.g.,][]{lee03}.

Third, it is likely that more sensitive observations could detect harmonic
content in the profiles of the burst
\begin{figure}[t]
\resizebox{.9\columnwidth}{!}
 {\includegraphics{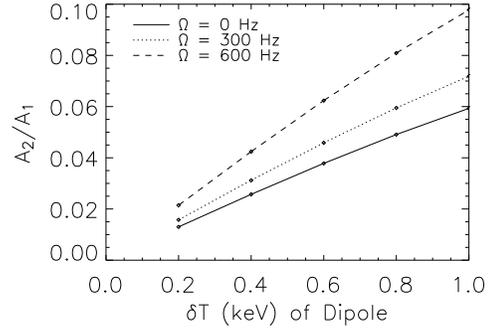}}
\caption{Plot of the ratios of the amplitude of the harmonic ($A_2$) to 
that of the fundamental signal ($A_1$) produced by a dipole ($\cos \theta$) 
temperature distribution of varying temperature contrast $\delta T$ and 
neutron star rotation $\Omega$. The amplitudes of the oscillations as a 
function of energy indicate that $\delta T \approx 0.2$ keV 
(Muno et al. 2003). The current upper limit on the ratio
$A_2/A_1$ is $0.5$\%. A future X-ray timing mission with 10 times 
the effective area could detect harmonics with a factor of 3 lower 
fractional amplitude.
\label{fig:dipole}}
\end{figure}
 oscillations. For instance, a 
dipolar ($\propto \cos\theta$) temperature distribution on a neutron star 
should produce oscillations that are slightly-non sinusoidal, because the 
flux then would be distributed as $\cos^4\theta$ (Figure~\ref{fig:dipole}). 
These harmonics would be
just below the detection threshold of {\it RXTE}, 
but would be easily detectable
with a timing mission with larger area. As is discussed
by Tod Strohmayer in these proceedings, the harmonic content of the
oscillations is crucial to constraining the compactness of the neutron star,
and hence its equation of state 
\citep[see also][]{ml98, wml01, brr00, nss02,bat04}.

Thus, future observations of millisecond oscillations during thermonuclear 
X-ray bursts could provide important insight into the physics of neutron 
stars.

\begin{theacknowledgments}
I would like to thank D. Chakrabarty, D. Fox, D. Galloway, J. Hartman, F.
\"{O}zel, and D. Psaltis for their significant contributions to the work I
have participated in on this topic. This review was written with support
from a Hubble Fellowship from the Space Telescope Science Institute, which is 
operated by the Association of Universities for Research in Astronomy, Inc., 
under NASA contract NAS 5-26555.
\end{theacknowledgments}

\end{document}